\documentclass[10pt, onecolumn, aps, pra, showkeys, longbibliography]{revtex4-2}

\usepackage{orcidlink}
\usepackage[T1]{fontenc}
\usepackage[utf8]{inputenc}
\usepackage[english]{babel}
\usepackage{braket}
\usepackage{xcolor}
\usepackage{natbib}
\usepackage{amsfonts, amsmath, amssymb}
\usepackage{graphicx}
\usepackage[top = 2cm, left = 2.5cm, right = 2.5cm, bottom = 2cm]{geometry}
\usepackage{hyperref}
\hypersetup{
    colorlinks = true,
    allcolors = blue,
    pdftitle = {},
}

\newcommand{\mi}{\mathrm{i}}

\begin{document}
\title{A Pedagogical Derivation of the First-Order Effective Hamiltonian for the Two-Mode Jaynes–Cummings Model}
\author{Alejandro R. Urz\'ua\,\orcidlink{0000-0002-6255-5453}}
\email{alejandro@icf.unam.mx}
\affiliation{
Instituto de Ciencias F\'isicas, Universidad Nacional Aut\'onoma de M\'exico\\ Av. Universidad sn, Col. Chamilpa, 62210, Cuernavaca, Morelos, M\'exico\\
Facultad de Ciencias Qu\'imicas e Ingeniería, Universidad Aut\'onoma del Estado de Morelos.\\ Av. Universidad 1001, Col. Chamilpa, 62210, Cuernavaca, Morelos, M\'exico}

\date{\today}

\begin{abstract}
    This work presents a pedagogical and self-contained derivation of the first-order effective Hamiltonian for the two-mode Jaynes–Cummings model in the dispersive regime. A perturbative unitary transformation removes nonresonant atom–field terms, revealing dispersive frequency shifts leading to an atom-induced effective beam-splitter interaction between the field modes. The resulting Hamiltonian is diagonalized through a simple geometric rotation in the two-mode bosonic space, providing a transparent interpretation of the underlying dynamics. The exposition emphasizes clarity and physical insight, making effective Hamiltonian methods accessible for teaching and learning in multimode light–matter interactions.
\end{abstract}

\keywords{two-mode Jaynes–Cummings model; dispersive regime; effective Hamiltonian; mode–mode interaction; beam-splitter coupling; multimode quantum optics; pedagogical quantum methods}

\maketitle

\section{Introduction}

The Jaynes--Cummings model (JCM) is one of the most familiar and instructive systems in which to study quantized light--matter interaction. Its original formulation by Jaynes and Cummings~\cite{JC1963} introduced a fully quantum description of a two-level atom interacting with a single mode of the electromagnetic field. Its dynamical features, like Rabi oscillations and collapse and revival, remain standard examples in textbooks and advanced courses in quantum optics. Comprehensive pedagogical expositions can be found in works such as Haroche and Raimond~\cite{HarocheRaimond} and related treatments that build upon the JCM to introduce coherent dynamics, cavity quantum electrodynamics (cavity QED), and atom--field entanglement \cite{Larson2024}. Modern experimental platforms greatly extend this basic picture. In cavity QED, trapped ions, and especially circuit QED, a single atomic or artificial atomic system often couples to more than one bosonic mode~\cite{BlaisRMP, WallraffNature, LeibfriedRMP, HuertaRodriguezLara2018}. In these settings, the JC Hamiltonian naturally generalizes to multimode versions in which the atom acts as a mediator between modes of different frequencies. Recent work has also considered extensions of the Jaynes–Cummings interaction to finite-dimensional oscillator systems, providing a framework for understanding one and two-mode interactions in constrained Hilbert spaces and their excitation-conserved eigenbases~\cite{MartinRuiz2014, Urzua2509.07215}. It is thus that the two-mode JCM is the simplest and most transparent example of this generalization, as it captures essential mechanisms underlying mode hybridization, conditional frequency shifts, and atom-induced correlations that appear across quantum optical and quantum information applications.

However, even this two-mode extension introduces conceptual and mathematical complications compared with the single-mode case. In many experimentally relevant regimes, the system operates far from resonance, and real excitation exchange between the atom and each mode is strongly suppressed. Instead, the dominant effects are dispersive, where the atom shifts the energies of the field modes and induces an effective interaction between them. Understanding these features requires moving beyond exact diagonalization and adopting approximation techniques that retain physical meaning while simplifying the dynamics. A particularly powerful tool in this matter is the \emph{effective Hamiltonian} method, in which a perturbative unitary transformation removes nonresonant terms order by order in the small parameter characterizing the dispersive regime. This approach, closely related to the Schrieffer--Wolff (SW) transformation widely used in quantum information theory and many-body physics, has been developed and reviewed in Refs.~\cite{BravyiSW, JamesJerke, Klimov2002}. In quantum optics, it provides an intuitive way to understand how virtual processes give rise to frequency shifts, nonlinear corrections, and mode--mode couplings. Despite its broad relevance, the effective Hamiltonian method is often introduced in ways that are technically efficient but not fully pedagogical. Intermediate steps are commonly omitted, and the physical meaning of the resulting terms, especially the beam-splitter type interaction that emerges in the two-mode case, is not always made explicit. As a result, students encountering multimodal light--matter interaction for the first time can find the technique opaque. In this work, a pedagogical and self-contained derivation of the first-order effective Hamiltonian for the two-mode Jaynes--Cummings model in the dispersive regime is presented. The aim is to give a clear illustration of how a perturbative unitary transformation eliminates the nonresonant atom--field terms and how the remaining interaction can be interpreted as an effective mixing between the field modes. A simple geometric rotation in the two-mode bosonic space then diagonalizes the resulting Hamiltonian and provides transparent physical insight into the atom-mediated coupling.

Throughout the exposition, clarity, physical intuition, and explicit intermediate steps are emphasized. This tutorial-style treatment is intended for students and instructors seeking a clear path from the basic JCM to its multimode extensions and to the effective Hamiltonian techniques that are now widely used in modern cavity, trapped-ion, and circuit QED platforms. The remainder of this work is organized as follows. Section~\ref{sec_2} reviews the single-mode Jaynes–Cummings model and the dispersive regime. Section~\ref{sec_3} introduces the two-mode Hamiltonian and the conditions under which a perturbative treatment applies. Section~\ref{sec_4} derives the first-order effective Hamiltonian via a small-rotation transformation, and Section~\ref{sec_5} diagonalizes it through a geometric rotation of the field modes. Section~\ref{sec_6} examines the resulting dynamics and illustrates the slow, atom-conditioned mode mixing characteristic of the dispersive limit. Section~\ref{sec_cons} summarizes the main insights and comments on extensions to broader multimode settings.

\section{Review of the JCM and the Dispersive Regime}
\label{sec_2}

Before addressing the two-mode JCM, it is helpful to recall the structure and physical interpretation of the standard single-mode case. This section summarizes the key features of the one-mode Jaynes--Cummings Hamiltonian, its dressed-state structure, and the conditions under which the system enters the dispersive regime. The presentation is intended as a brief but self-contained review to establish the notation and intuition used in later sections.

\subsection{The one-mode JCM Hamiltonian}

A two-level atom with a ground state $\lvert -\rangle$, an excited state $\lvert +\rangle$,
and a transition frequency $\Omega_{0}$ interacting with a single quantized field mode of
frequency $\omega$ is described by the JCM Hamiltonian~\cite{JC1963},
\begin{equation}
\label{H_JC_single}
    \hat{\mathcal{H}}_{\texttt{JCM}} = \omega\hat{a}^{\dagger}\hat{a} + \Omega_{0}\hat{\sigma}_{z} + g\left(\hat{a}^{\dagger}\hat{\sigma}_{-} + \hat{a}\hat{\sigma}_{+}\right),
\end{equation}
where $\hat{a}$ and $\hat{a}^{\dagger}$ are the annihilation and creation operators of the field mode, and $g$ is the atom--field coupling strength. The atomic operators $\hat{\sigma}_{\pm}$ and $\hat{\sigma}_{z}$ are the usual Pauli ladder and inversion operators.

A defining feature of the JCM Hamiltonian is the conservation of the total excitation number,
\begin{equation}
    \hat{\mathcal{N}} = \hat{a}^{\dagger}\hat{a} + \hat{\sigma}_{+} \hat{\sigma}_{-},
\end{equation}
which allows the dynamics to be solved independently in each excitation subspace. Diagonalizing $\hat{\mathcal{H}}_{\texttt{JCM}}$ therefore yields \emph{dressed states} (also known as polaritons or hybrid light-matter quasiparticles) in each excitation manifold and produces the well-known Rabi splitting $\pm\sqrt{\Delta^{2} + 4g^2(n+1)}$, where $\Delta = \Omega_{0}-\omega$ is the atom--field detuning.

These results, discussed extensively in Refs.~\cite{HarocheRaimond}, provide intuition for how quantized light can exchange excitations with a two-level system and illustrate the nonclassical phenomena that arise from coherent atom--field coupling.

\subsection{Large detuning and the dispersive limit}

When the detuning $\Delta$ is large compared with the effective coupling strength,
\begin{equation}
\label{dispersive_cond}
    \vert\Delta\vert \gg g \sqrt{\langle\hat{n}\rangle + 1},
\end{equation}
the atom and the field cannot efficiently exchange real excitations. Instead, their interaction is dominated by \emph{virtual} processes where the atom is only briefly excited, and the photon number remains essentially unchanged during the evolution. In this limit, the JCM Hamiltonian may be approximated by an effective Hamiltonian in which excitation exchange terms are eliminated perturbatively.

Keeping terms up-to second order in $g/\Delta$, one finds the familiar dispersive
Hamiltonian,
\begin{equation}
\label{H_disp_single}
    \hat{\mathcal{H}}_{\texttt{disp}} = \omega\hat{a}^\dagger\hat{a} + \Omega_{0}\hat{\sigma}_{z} + \chi\hat{a}^\dagger\hat{a}\hat{\sigma}_{z},
\end{equation}
where $\chi = g^{2}/\Delta$ is the dispersive shift. This result can be obtained through various perturbative methods, including the Schrieffer--Wolff transformation, widely used in many-body physics~\cite{BravyiSW, JamesJerke}. Going deep into what equation~(\ref{H_disp_single}) reveals, we identify two key features of the dispersive regime. First, the field frequency is shifted depending on the atomic state $\omega \rightarrow \omega\pm\chi$. Second, the atomic transition frequency acquires a photon-number-dependent Stark shift $\Omega_{0} \rightarrow \Omega_{0} + 2\chi \langle n\rangle.$ Both corrections arise from virtual processes in which the atom is excited momentarily and then returns to its original state. The dispersive Hamiltonian thus captures the leading effects of the JCM interaction when real energy exchange is suppressed.

The dispersive regime plays a central role in modern quantum technologies. In circuit QED, for example, the qubit is routinely operated in the dispersive regime to implement quantum nondemolition readout of cavity photon number or qubit state~\cite{BlaisRMP, WallraffNature}. In trapped-ion systems, dispersive couplings underlie common schemes for quantum gates and state preparation~\cite{LeibfriedRMP}. From an educational perspective, the dispersive approximation provides an accessible introduction to effective Hamiltonian techniques. It makes clear how virtual processes shape the physics, and how perturbative unitary transformations can simplify dynamics by eliminating nonresonant terms. These ideas extend naturally to multimode systems, where the atom can mediate interactions between different bosonic modes and generate new effective couplings that do not appear in the single-mode case. In the next section, we extend this reasoning to the two-mode Jaynes--Cummings model, showing how the same perturbative logic leads to an effective Hamiltonian that contains not only dispersive shifts but also an atom-induced coupling between the modes. This example provides a concrete and physically intuitive setting in which to understand the structure and usefulness of effective Hamiltonians in multimode light--matter interactions.

\section{The Two-Mode JCM Hamiltonian}
\label{sec_3}

We now extend the discussion from the single-mode case to the situation in which a two-level atom interacts simultaneously with two distinct quantized field modes. This is the simplest multimode generalization of the Jaynes--Cummings model and provides a clear setting in which to study atom mediated interactions between bosonic modes.

\subsection{Hamiltonian and physical setting}

Let the atom couple to two independent harmonic oscillators of frequencies $\omega_{a}$ and $\omega_{b}$. The corresponding creation and annihilation operators are $\hat{a}^{\dagger} (\hat{a})$ and $\hat{b}^{\dagger} (\hat{b})$, and the atom--field independent coupling strengths for each mode are $g_{a}$ and $g_{b}$. The full Hamiltonian is given then by
\begin{equation}
\label{H_2mode_full}
    \hat{\mathcal{H}}_{\texttt{2JCM}}
    = \omega_{a} \hat{a}^{\dagger} \hat{a}
    + \omega_{b} \hat{b}^{\dagger} \hat{b}
    + \Omega_{0} \hat{\sigma}_{z}
    + g_{a} \left(\hat{a}^{\dagger} \hat{\sigma}_{-} + \hat{a}\hat{\sigma}_{+}\right)
    + g_{b} \left(\hat{b}^{\dagger} \hat{\sigma}_{-} + \hat{b}\hat{\sigma}_{+}\right),
\end{equation} 
where the first two terms represent the free evolution of the field modes, the third term is the atomic energy, and the last two terms describe the atom exchanging excitations with each mode. This Hamiltonian has the same structure as the single-mode JCM Hamiltonian, but now the atom serves as an intermediary between two otherwise independent oscillators. When $g_{a}$ and $g_{b}$ are both comparable, the atom can transfer energy between the modes or imprint correlations on their joint state, even though the modes do not interact directly.

As in the single-mode case, the atom--field interaction conserves the \emph{total} excitation number,
\begin{equation}
    \hat{\mathcal{N}}
    = \hat{a}^\dagger \hat{a}
    + \hat{b}^\dagger \hat{b}
    + \hat{\sigma}_{+} \hat{\sigma}_{-},
\end{equation}
but not the individual photon numbers $\hat{n}_{a} = \hat{a}^\dagger\hat{a}$ and $\hat{n}_{b} = \hat{b}^{\dagger}\hat{b}$. Within each $\hat{N}$-subspace, the Hamiltonian is block diagonal and couples states such as
\begin{equation}
    \ket{n_{a}, n_{b}, g} \leftrightarrow \ket{n_{a} + 1, n_{b}, e},\qquad
    \ket{n_{a}, n_{b}, g} \leftrightarrow \ket{n_{a}, n_{b} + 1, e}
\end{equation}
where the resulting dynamics are richer than in the single-mode model. Even at resonance, the two couplings $g_{a}$ and $g_{b}$ compete, giving rise to hybridization between sectors in which the atom exchanges excitations with mode $a$ or with mode $b$. Analytic diagonalization of the full Hamiltonian is therefore considerably more convoluted. For pedagogical purposes, this makes the model an ideal candidate for illustrating how effective Hamiltonians can simplify complex multimode dynamics.

To connect with the effective Hamiltonian approach, we introduce the detunings
\begin{equation}
    \Delta_{a} = \Omega_{0} - \omega_{a},
    \qquad
    \Delta_{b} = \Omega_{0} - \omega_{b},
\end{equation}
where the focus is on the regime in which \emph{both} couplings are far from resonance, this is
\begin{equation}
\label{dispersive_2mode_condition}
    \vert\Delta_{a}\vert \gg g_{a} \sqrt{\langle \hat{n}_{a} \rangle + 1},
    \qquad
    \vert\Delta_{b}\vert \gg g_{b} \sqrt{\langle \hat{n}_b \rangle + 1},
\end{equation}
so that real excitation exchange between the atom and the field modes is strongly suppressed. In this limit, the dominant effects of the interaction arise from virtual processes as in the one-mode case. Each mode experiences a dispersive frequency shift proportional to $g_{k}^2/\Delta_{k}$ for $k\in\{a, b\}$, and, crucially, the two modes become \emph{effectively coupled} to each other through the atom. Such atom-mediated interactions are important in several modern settings. In circuit QED, for example, two microwave resonators can be coupled by a single qubit acting as a tunable quantum bus; in optical systems, a single atom in a multimode cavity can induce coherent mode mixing. In all these cases, the effective Hamiltonian framework provides a transparent way to identify and quantify the resulting interactions.

At first, the two-mode JCM Hamiltonian in Eq.~\eqref{H_2mode_full} looks simple, but its exact solution does can be cumbersome due to its convolved algebraic structure. The dispersive regime, however, suggests a clear simplification if excitation exchange is weak; one may eliminate the nonresonant terms perturbatively and obtain a Hamiltonian that captures the leading effects---frequency shifts and mode--mode mixing---compactly and intuitively. In the next section, we construct such an effective Hamiltonian using a small-angle unitary transformation. This procedure extends the logic of the single-mode dispersive conversion and reveals how the atom generates an effective beam-splitter interaction between the modes, even though no direct coupling is present in the original Hamiltonian.

\section{Unitary Transformation and the First-Order Effective Hamiltonian}
\label{sec_4}

In the dispersive regime introduced in the previous section, real exchange of excitations between the atom and each field mode is strongly suppressed. Nevertheless, the interaction terms in Eq.~\eqref{H_2mode_full} still influence the dynamics through virtual processes. A standard way to isolate these effects and eliminate the nonresonant excitation-exchange terms is to apply a perturbative unitary transformation. This method, sometimes referred to as a ``small rotation'' or SC transformation, has been widely used in quantum optics and condensed-matter physics.

Consider the interaction part of the Hamiltonian,
\begin{equation}
\label{V_int_2mode}
    \hat{\mathcal{V}}_{\texttt{int}}
    = g_{a} \left(\hat{a}^\dagger\hat{\sigma}_{-} + \hat{a}\hat{\sigma}_{+}\right)
    + g_b \left(\hat{b}^{\dagger}\hat{\sigma}_{-} + \hat{b}\hat{\sigma}_{+}\right),
\end{equation}
which contains the nonresonant excitation-exchange operators. When $\vert\Delta_a\vert$ and $\vert\Delta_b\vert$ are large, the effect of these operators can be removed order by order in a suitable small parameter regime. This is accomplished by choosing a unitary operator of the form
\begin{equation}
\label{S_unitary}
    \hat{\mathcal{S}}(\epsilon_a,\epsilon_b)
    = \exp\!\left\{
        \epsilon_{a}(\hat{a}\hat{\sigma}_{+} - \hat{a}^\dagger\hat{\sigma}_{-})
        + \epsilon_{b}(\hat{b}\hat{\sigma}_{+} - \hat{b}^\dagger\hat{\sigma}_{-})
    \right\},
\end{equation}
with $\epsilon_{a}$ and $\epsilon_{b}$ small parameters to be determined. This structure is motivated by perturbative diagonalization techniques described in Refs.~\cite{CohenTannoudjiAPI,SchleichQO} and by the general form of the Schrieffer--Wolff method~\cite{BravyiSW,JamesJerke}. In the first-order expansion in $\epsilon_{a}$ and $\epsilon_{b}$, a suitable Hamiltonian is transformed as
\begin{equation}
\begin{aligned}
\label{Hprime_firstorder}
    &\hat{\mathcal{H}}' = \hat{\mathcal{S}}^{\dagger} \hat{\mathcal{H}} \hat{\mathcal{S}}
    \approx \hat{\mathcal{H}}
    + [\hat{\mathcal{H}}, \hat{G}]
    + \mathcal{O}(\epsilon_{a}^{2}, \epsilon_{b}^{2}),\\
    &
    \text{with}\quad
    \hat{G} \equiv
    \epsilon_{a}(\hat{a}\hat{\sigma}_{+} - \hat{a}^{\dagger}\hat{\sigma}_{-})
    + \epsilon_{b}(\hat{b}\hat{\sigma}_{+} - \hat{b}^{\dagger}\hat{\sigma}_{-}).
\end{aligned}
\end{equation}
Our goal is to choose $\epsilon_{a}$ and $\epsilon_{b}$ such that all terms of the form $\hat{a}\hat{\sigma}_{+}$, $\hat{a}^\dagger\hat{\sigma}_{-}$, $\hat{b}\hat{\sigma}_{+}$, and $\hat{b}^\dagger\hat{\sigma}_{-}$ cancel up-to first order.

Let
\begin{equation}
    \hat{\mathcal{H}}_{0} = \omega_{a} \hat{a}^{\dagger}\hat{a}
    + \omega_{b}\hat{b}^{\dagger}\hat{b}
    + \Omega_{0}\hat{\sigma}_{z},
\end{equation}
denote the unperturbed Hamiltonian. A direct calculation shows that
\begin{equation}
    [\hat{\mathcal{H}}_{0}, \hat{a}^\dagger\hat{\sigma}_{-}]
    = -\Delta_{a}\hat{a}^{\dagger}\hat{\sigma}_{-},\qquad
    [\hat{\mathcal{H}}_{0}, \hat{a}\hat{\sigma}_{+}]
    = +\Delta_{a}\hat{a}\hat{\sigma}_{+},
\end{equation}
and similarly for mode $b$ with $\Delta_{b}$. Using these relations, the first-order transformed Hamiltonian becomes
\begin{equation}
\begin{aligned}
    \hat{\mathcal{H}}' &\approx \hat{\mathcal{H}}_{0} + \hat{V}_{\texttt{int}}
    + [\hat{\mathcal{H}}_{0}, \hat{G}] \\
    &= \hat{\mathcal{H}}_0
    + (g_{a} - \epsilon_{a}\Delta_{a})(\hat{a}^{\dagger} \hat{\sigma}_{-}
    + \hat{a}\hat{\sigma}_{+})
    + (g_{b} - \epsilon_{b}\Delta_b)(\hat{b}^{\dagger} \hat{\sigma}_{-}
    + \hat{b}\hat{\sigma}_{+}) + \mathcal{O}(\epsilon_{a}^{2}, \epsilon_{b}^{2}).
\end{aligned}
\end{equation}
where we can conclude that the nonresonant terms vanish to first order if
\begin{equation}
\label{epsilons}
    \epsilon_{a} = \frac{g_{a}}{\Delta_{a}},
    \qquad
    \epsilon_{b} = \frac{g_{b}}{\Delta_{b}}.
\end{equation}

With the choices in Eq.~\eqref{epsilons}, all terms linear in the excitation-exchange operators cancel at order $g_{k}/\Delta_{k}$ for $k\in\{a, b\}$. The remaining Hamiltonian contains only diagonal terms in the atomic basis. Keeping the leading second-order contributions, one obtains
\begin{equation}
\label{Heff_firstorder}
    \hat{\mathcal{H}}_{\texttt{eff}}
    = \omega_{a}\hat{a}^{\dagger}\hat{a}
    + \omega_{b}\hat{b}^{\dagger}\hat{b}
    + \Omega_{0}\hat{\sigma}_{z}
    + \frac{2g_{a}^{2}}{\Delta_{a}}
      \hat{a}^{\dagger}\hat{a}\hat{\sigma}_{z}
    + \frac{2g_{b}^2}{\Delta_b}
      \hat{b}^\dagger\hat{b}\hat{\sigma}_{z}
    + \frac{2g_{a}g_{b}}{\Delta_{ab}}
      \left(\hat{a}^\dagger \hat{b} + \hat{a}\hat{b}^\dagger\right)\hat{\sigma}_{z},
\end{equation}
where
\begin{equation}
    \frac{1}{\Delta_{ab}}
    = \frac{1}{\Delta_{a}} + \frac{1}{\Delta_{b}},
\end{equation}
which is not a physical detuning but a convenient shorthand encoding of the two virtual excitation paths.

The first two dispersive terms, $(2g_{a}^{2}/\Delta_{a})\hat{a}^\dagger\hat{a}\hat{\sigma}_{z}$ and $(2g_{b}^{2}/\Delta_{b})\hat{b}^\dagger\hat{b}\hat{\sigma}_{z}$, represent the usual AC Stark shifts of the field modes conditioned on the atomic state. The third term,
\begin{equation}
\label{eq:mixing_modes}
    \frac{2g_{a}g_{b}}{\Delta_{ab}}
    \left(\hat{a}^\dagger \hat{b} + \hat{a}\hat{b}^\dagger\right)\hat{\sigma}_{z},
\end{equation}
is an \emph{effective beam-splitter interaction} between the two modes, mediated entirely by virtual excitations of the atom. This term has no counterpart in the single-mode JC model and is the clearest signature of multimode physics in the dispersive limit. Its presence illustrates how effective Hamiltonian techniques reveal new emergent interactions even when the original Hamiltonian contains no direct mode--mode coupling. In the next section, we show how this first-order effective Hamiltonian can be diagonalized by a simple geometric rotation in the two-mode bosonic space, yielding a transparent interpretation of the atom-induced mixing between modes.

\section{Geometric Diagonalization of the Effective Hamiltonian}
\label{sec_5}

The first-order effective Hamiltonian derived in Eq.~\eqref{Heff_firstorder} contains three diagonal dispersive contributions and a single off-diagonal term, which mixes the two bosonic modes. This operator is the generator of a $\mathfrak{su}(2)$ algebra, which lead to a beam-splitter transformation~\cite{CamposSalehTeich1989, GerryKnight}, and its presence indicates that the atom mediates a coherent mode–mode coupling even though the original Hamiltonian contains no direct interaction between $\hat{a}$ and $\hat{b}$. A key operational and pedagogical advantage of the effective-Hamiltonian method is that once all nonresonant excitation-exchange terms have been removed, the remaining quadratic Hamiltonian can be diagonalized by a simple geometric rotation in the two-dimensional bosonic mode space. In this section, we construct that rotation and show how it leads to two normal modes whose frequencies depend explicitly on the atomic state.

Throughout this section, we restrict attention to atomic energy eigenstates; atomic superpositions would entangle the atom with the rotated field modes and require a block diagonal rather than scalar rotation. Because the effective Hamiltonian is diagonal in $\hat{\sigma}_{z}$, we may work in the atomic eigenbasis $\hat{\sigma}_{z}\ket{\pm} = \pm\ket{\pm}$. In each manifold, the Hamiltonian reduces to
\begin{equation}
\label{eq:Heff_s}
    \hat{\mathcal{H}}_{\texttt{eff}}^{(\pm)}
    =
    \tilde{\omega}_{a}^{(\pm)}\hat{a}^{\dagger}\hat{a}
    +
    \tilde{\omega}_{b}^{(\pm)}\hat{b}^{\dagger}\hat{b}
    \pm
    J~\left(\hat{a}^{\dagger}\hat{b}+\hat{a}\hat{b}^{\dagger}\right),
\end{equation}
where $J = \tfrac{2g_{a}g_{b}}{\Delta_{ab}}$, and with dispersively shifted frequencies given by
\begin{equation}
    \tilde{\omega}_a^{(\pm)}=\omega_{a} \pm \frac{2g_{a}^2}{\Delta_a},
    \qquad
    \tilde{\omega}_b^{(\pm)}=\omega_{b} \pm \frac{2g_b^2}{\Delta_b}.
\end{equation}
Thus, in each atomic subspace, diagonalization reduces to solving a two-mode beam-splitter Hamiltonian—an $\mathfrak{su}(2)$ rotation problem. Diagonalizing the effective Hamiltonian is equivalent to choosing the optical basis in which the atom-induced beam splitter is aligned. Moreover, from this perspective, the atom functions as a state-dependent linear optical element, selecting the effective beam-splitter angle and phase.

The standard rotation operator for two bosonic modes is~\cite{CamposSalehTeich1989, BarnettRadmore}
\begin{equation}
\label{eq:Rtheta}
    \hat{\mathcal{R}}(\theta)
    =
    \exp\!\left[\theta\left(\hat{a}^{\dagger}\hat{b} - \hat{a}\hat{b}^{\dagger}\right)\right],
\end{equation}
which transforms the modes according to
\begin{equation}
    \hat{A} = \hat{\mathcal{R}}^{\dagger}(\theta)\hat{a}\hat{\mathcal{R}}(\theta)
        = \hat{a}\cos\theta - \hat{b}\sin\theta,\qquad
    \hat{B} = \hat{\mathcal{R}}^\dagger(\theta)\hat{b}\hat{\mathcal{R}}(\theta)
        = \hat{b}\cos\theta + \hat{a}\sin\theta. 
\end{equation}

Applying $\hat{\mathcal{R}}(\theta)$ to Eq.~\eqref{eq:Heff_s}, the Hamiltonian includes a term proportional to $(\hat{A}^\dagger\hat{B} + \hat{A}\hat{B}^\dagger)$. Requiring its coefficient to vanish yields the diagonalization condition
\begin{equation}
\label{eq:theta_s}
    \tan\!\left(2\theta_\pm\right)
    =
    \frac{\pm 2J}{\tilde{\omega}_{a}^{(\pm)} - \tilde{\omega}_{b}^{(\pm)}}
    =
    \frac{\pm 2J}{(\omega_{a} - \omega_{b})
                 \pm\!\left(\frac{2g_{a}^{2}}{\Delta_{a}} - \frac{2g_{b}^{2}}{\Delta_{b}}\right)}.
\end{equation}

\begin{figure}[htbp]
\centering
    \includegraphics[width = \linewidth]{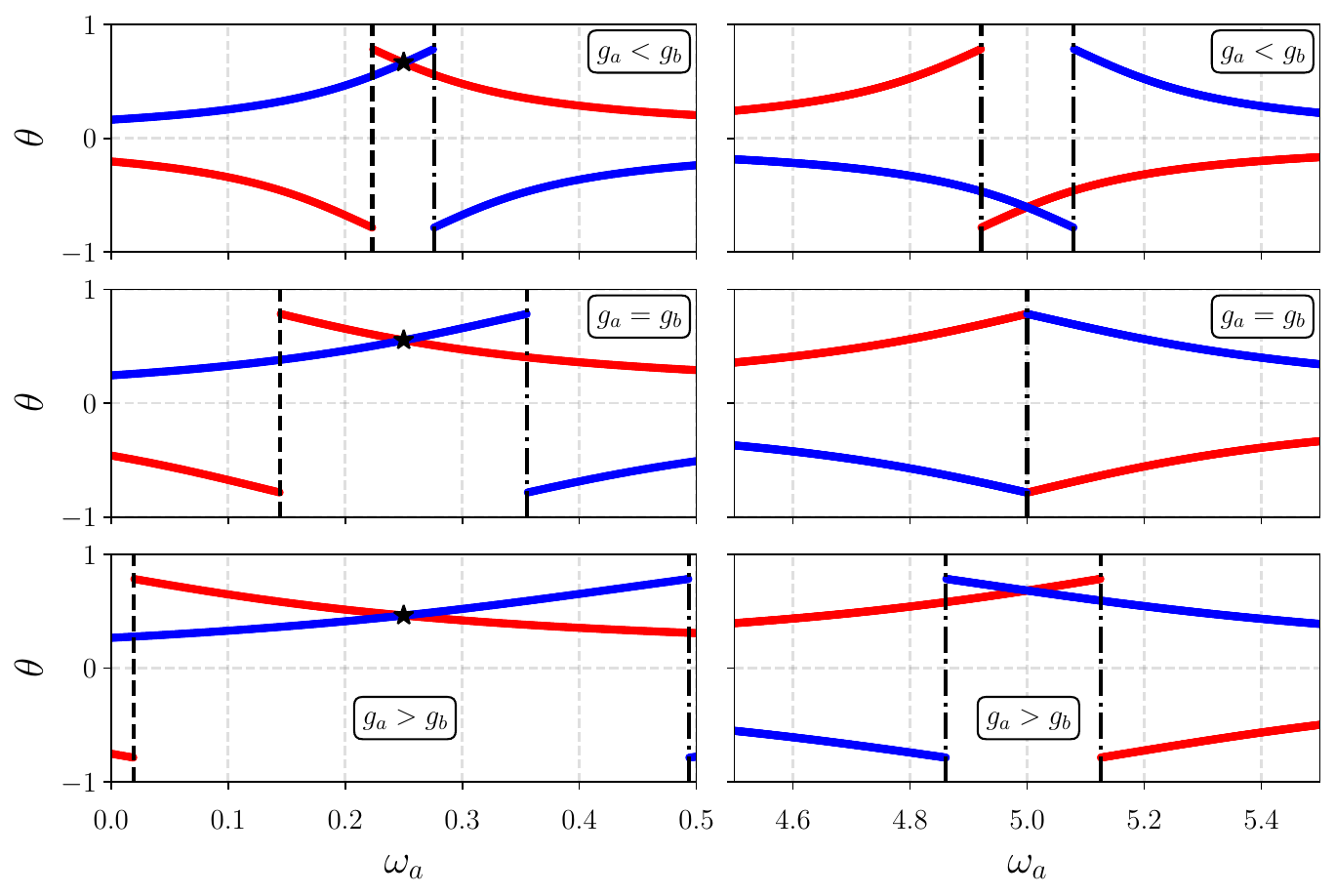}
    \caption{Branches of the rotation angle $\theta_\pm$ obtained from the diagonalization condition in Eq.~\eqref{eq:theta_s}. The two branches correspond to the atomic eigenstates $\ket{\pm}$ and represent the angles required to rotate the original fieldmodes $(\hat{a},\hat{b})$ into the normal modes $(\hat{A},\hat{B})$ that diagonalize the effective Hamiltonian in each atomic subspace. The vertical asymptotes occur when the effective detuning $\tilde{\omega}_a^{(s)}-\tilde{\omega}_b^{(s)}$ approaches zero, signaling near-degeneracy of the dispersively shifted modes. In this regime, a small change in parameters produces a rapid variation of $\theta_s$, reflecting strong atom-mediated mode hybridization. Away from these points, $\theta_s$ varies smoothly, and the normal modes remain predominantly aligned with one of the original field modes. This figure provides a geometric interpretation of the atom’s role as a tunable media of mode mixing: the atomic state selects a specific rotation branch, thereby controlling how strongly the two modes are hybridized in the dispersive regime.}
    \label{fig:theta_branches}
\end{figure}

This expression shows first that the two branches $\theta_{+}$ and $\theta_{-}$ correspond to the atomic states $\ket{\pm}$, and second that the rotation angle may vary strongly when the denominator approaches zero, corresponding to an avoided crossing of the mode frequencies. These features are precisely what appear in the numerical sweeps shown in Fig.~\ref{fig:theta_branches}.  There, the red and blue curves represent the two atomic branches $\theta_{\pm}$, while the vertical asymptotes identify parameter values for which the effective detuning $\tilde{\omega}_{a}^{(\pm)} - \tilde{\omega}_{b}^{(\pm)}$ crosses zero. The abrupt sign changes visible near these asymptotes reflect the rapid rotation required to diagonalize the Hamiltonian when the two dispersively shifted modes become nearly degenerate. This graphical behavior provides a clear geometric intuition for the role of the atom as a tunable mediator of mode mixing.

With $\theta = \theta_{\pm}$, the Hamiltonian becomes fully diagonal, as we can see
\begin{equation}
\label{eq:Heff_diag}
    \hat{\mathcal{H}}_{\texttt{eff}}^{(\pm)}
    =
    \Omega_{A}^{(\pm)}\hat{A}^{\dagger}\hat{A}
    +
    \Omega_{B}^{(\pm)}\hat{B}^{\dagger}\hat{B}
    +
    \text{const.},
\end{equation}
where the normal-mode frequencies are
\begin{equation}
\label{eq:Omega_AB}
    \Omega_{A, B}^{(\pm)}
    =
    \frac{\tilde{\omega}_{a}^{(\pm)} + \tilde{\omega}_{b}^{(\pm)}}{2}
    \pm
    \frac{1}{2}
    \sqrt{
        \left(\tilde{\omega}_{a}^{(\pm)} - \tilde{\omega}_{b}^{(\pm)}\right)^{2}
        + 4J^{2}
    }.
\end{equation}

This result indicates that the atom induces both \emph{dispersive shifts} (through $\chi_a,\chi_b$) and \emph{mode hybridization} (through $J$), the normal-mode splitting depends sensitively on the atomic state, consistent with the distinct branches $\theta_{\pm}$ in Fig.~\ref{fig:theta_branches}. The rotation angle encapsulates how the system interpolates smoothly between dominantly $a$ or $b$ modes as parameters vary. This geometric picture will serve as the foundation for studying the time evolution of field observables in the next section.

\section{Dynamical evolution}
\label{sec_6}

\begin{figure}[htbp]
    \centering
    \includegraphics[width = \linewidth]{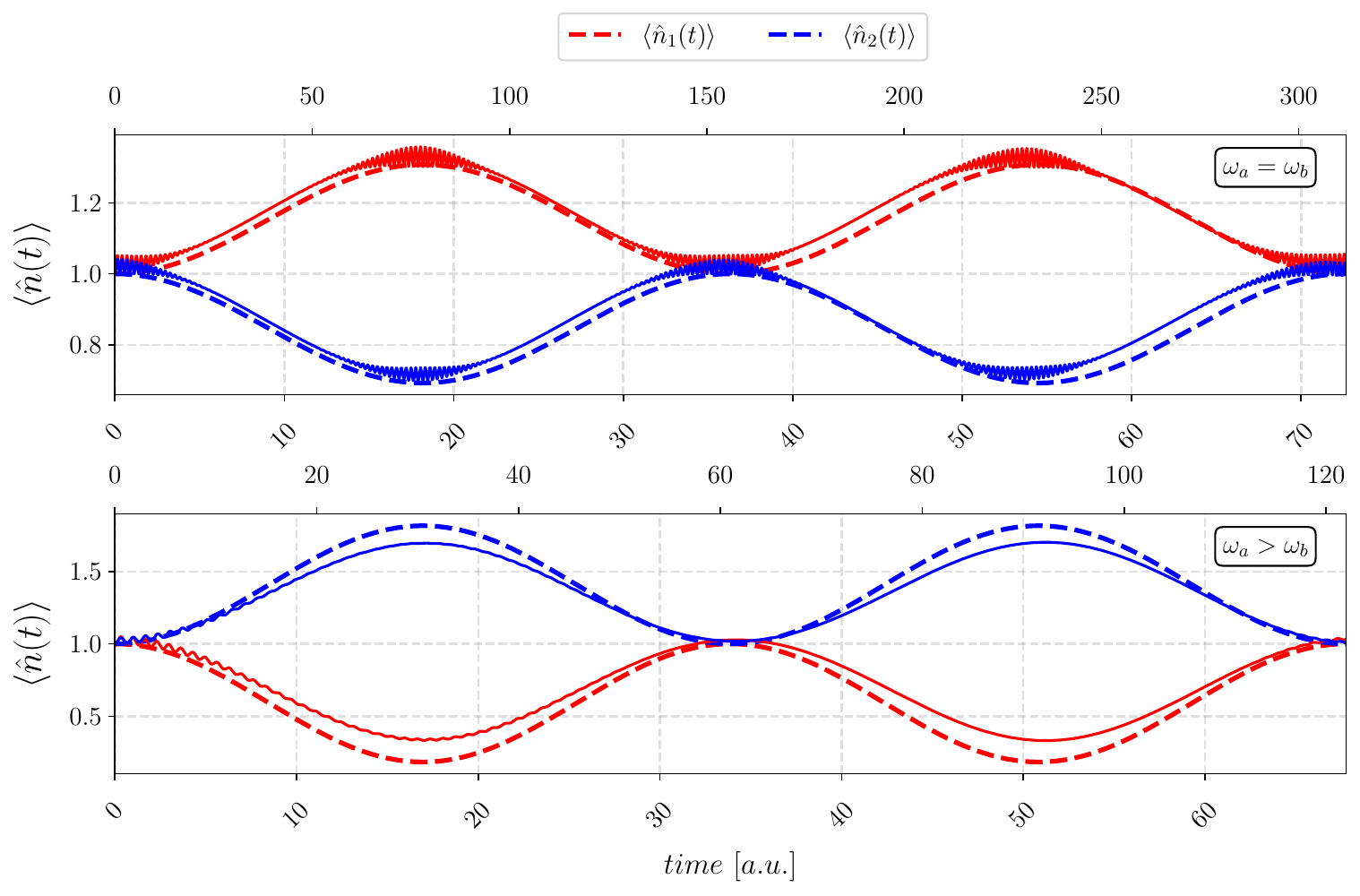}
    \caption{Time evolution of the mean photon numbers $\langle\hat{n}_a(t)\rangle$ and $\langle\hat{n}_b(t)\rangle$ for representative initial states and parameters, computed using the first-order effective Hamiltonian. Distinct curves correspond to the two atomic eigenstates $\ket{+}$, illustrating how the atomic state conditionally controls the field dynamics even though no real atom--field excitation exchange occurs in the dispersive regime. The oscillatory exchange of population between the modes is governed by the effective beam-splitter interaction induced by virtual atomic transitions. The characteristic oscillation period is set by the normal-mode splitting $|\Omega_{A}^{(\pm)} - \Omega_{B}^{(\pm)}|$, which scales as $g_{a} g_{b}/\Delta$ and is therefore much longer than the bare Rabi period $1/g_{k}$. This separation of time scales explains the slow dynamics observed (horizontal tick labels) against the normal time dynamics of the full system (tilded tick labels) in the figure, and illustrates how the effective-Hamiltonian treatment rescales the relevant time axis by eliminating fast atom--field oscillations. Overall, the figure demonstrates how the effective Hamiltonian reorganizes the dynamics into slow, physically transparent mode-exchange processes, whose frequency and amplitude are directly linked to the rotation angle $\theta_s$ and to the atomic state.}
    \label{fig:time_evolution}
\end{figure}

Once the effective Hamiltonian has been diagonalized in terms of the normal modes $\hat{A}$ and $\hat{B}$, the dynamics of the two-mode JCM become particularly transparent. In this section, the time evolution generated by the effective Hamiltonian is addressed, with worked examples for Fock and coherent initial states, and whose results permit an interpretation of the population dynamics shown in Fig.~\ref{fig:time_evolution}. Special emphasis is placed on understanding how the effective-Hamiltonian description introduces a \emph{rescaling of the relevant time scales} compared with the full microscopic dynamics.

In each atomic subspace, the effective Hamiltonian takes the diagonal form in Eq.~\eqref{eq:Heff_diag}; the corresponding time-evolution operator is given by
\begin{equation}
\label{eq:Ueff_time}
    \hat{\mathcal{U}}_{\texttt{eff}}^{(\pm)}(t)
    =
    \exp\!\left[
    -\mi~t\left(
    \Omega_{A}^{(\pm)}\hat{A}^{\dagger}\hat{A}
    +
    \Omega_{B}^{(\pm)}\hat{B}^{\dagger}\hat{B}
    \right)
    \right].
\end{equation}

Where a crucial point is that the characteristic frequencies governing the dynamics are no longer given in terms of the bare atom--field couplings $g_{a}$ and $g_{b}$, but rather to the \emph{second-order} energy scales
\begin{equation}
    \chi_{k} = \frac{2g_{k}^2}{\Delta_{k}},\quad k\in\{a, b\},
    \qquad
    J = \frac{2g_{a}g_{b}}{\Delta_{ab}},
\end{equation}
that arise from the virtual processes discussed above. As a result, the natural time scale of the dynamics is
\begin{equation}
    \tau_{\texttt{eff}} \sim \frac{1}{\big\vert\Omega_{A}^{(\pm)} - \Omega_{B}^{(\pm)}\big\vert},
\end{equation}
which is parametrically \emph{longer} than the microscopic Rabi period $1/g_{k}$. This separation of time scales is one of the key physical consequences of the dispersive approximation and explains why effective-Hamiltonian dynamics appear ``slowed down'' in time when compared with the full model.

Now, using the inverse rotation derived in Section~V,
\begin{equation}
    \hat{a} = \hat{A}\cos\theta_{\pm} + \hat{B}\sin\theta_{\pm},
    \qquad
    \hat{b} = -\hat{A}\sin\theta_{\pm} + \hat{B}\cos\theta_{\pm},
\end{equation}
the Heisenberg-picture operators evolve as
\begin{equation}
    \hat{a}(t) = \hat{\mathcal{U}}_{\texttt{eff}}^{(\pm)}{}^{\dagger}(t)~\hat{a}~\hat{\mathcal{U}}_{\texttt{eff}}^{(\pm)}(t) =
    \hat{a}~\mathcal{F}_{1}^{(s)}(t) + \hat{b}~\mathcal{F}_{2}^{(s)}(t),
\end{equation}
\begin{equation}
    \hat{b}(t) = \hat{\mathcal{U}}_{\texttt{eff}}^{(\pm)}{}^{\dagger}(t)~\hat{b}~\hat{\mathcal{U}}_{\texttt{eff}}^{(\pm)}(t) =
    \hat{a}~\mathcal{G}_{1}^{(s)}(t) + \hat{b}~\mathcal{G}_{2}^{(s)}(t),
\end{equation}
where the companion coefficients for $a(t)$ are given by
\begin{equation}
\begin{aligned}
    \mathcal{F}_{1}^{(\pm)}(t)
    &=
    \cos^{2}\theta_\pm\,e^{-\mi\Omega_{A}^{(\pm)}t}
    +
    \sin^{2}\theta_\pm\,e^{-\mi\Omega_{B}^{(\pm)}t},\\
    \mathcal{F}_{2}^{(\pm)}(t)
    &=
    \cos\theta_\pm\sin\theta_\pm
    \left(
    e^{-\mi\Omega_{A}^{(\pm)}t}
    -
    e^{-\mi\Omega_{B}^{(\pm)}t}
    \right),
\end{aligned}
\end{equation}
and analogous expressions for $\mathcal{G}_{1}^{(\pm)}(t)$ and $\mathcal{G}_{2}^{(\pm)}(t)$,
\begin{equation}
\begin{aligned}
    \mathcal{G}_{1}^{(\pm)}(t) &= e^{\mi\Omega_{B}^{(\pm)}t}\cos^{2}\theta_{\pm} + e^{\mi\Omega_{A}^{(\pm)}t}\sin^{2}\theta_{\pm}\\
        \mathcal{G}_{2}^{(\pm)}(t) &= \mathcal{F}_{2}^{(\pm)}(t).
\end{aligned}
\end{equation}

These expressions show explicitly that the dynamics arise from \emph{interference between two slow frequencies}, $\Omega_{A}^{(\pm)}$ and $\Omega_{B}^{(\pm)}$, rather than from fast atom--field oscillations.

\paragraph{Worked examples.} Set, for example, a general initial wavefunction $\ket{\psi(0)} = \ket{n_{a}, n_{b}, +}$, the action of the time-dependent operators given by the relations above, gives
\begin{equation}
    \begin{aligned}
        \hat{n}_{a}(t)&\ket{n, m, +} = \hat{a}^{\dagger}(t)\hat{a}(t)\ket{n, m, +}\\
        &= n\left\vert\mathcal{F}_{1}(t)\right\vert^{2}\ket{n, m, +} + m\left\vert\mathcal{F}_{2}(t)\right\vert^{2}\ket{n, m, +}\\
        & + \sqrt{(n + 1) m}\ \mathcal{F}_{1}(t)\mathcal{F}_{2}(t)^{*}\ket{n + 1, m - 1, +}
        + \sqrt{n (m + 1)}\ \mathcal{F}_{1}(t)^{*}\mathcal{F}_{2}(t)\ket{n - 1, m + 1, +},\\
        \hat{n}_{b}(t)&\ket{n, m, +} = \hat{b}^{\dagger}(t)\hat{b}(t)\ket{n, m, +}\\
        &= m\left\vert\mathcal{G}_{1}(t)\right\vert^{2}\ket{n, m, +} + n\left\vert\mathcal{G}_{2}(t)\right\vert^{2}\ket{n, m, +}\\
        & + \sqrt{(n + 1) m}\ \mathcal{G}_{1}(t)\mathcal{G}_{2}(t)^{*}\ket{n + 1, m - 1, +}
        + \sqrt{n (m + 1)}\ \mathcal{G}_{1}(t)^{*}\mathcal{G}_{2}(t)\ket{n - 1, m + 1, +},
    \end{aligned}    
\end{equation}
where, for simplicity, all the expressions required to be evaluated at $(\pm)\mapsto (+)$, carry no superscript, meaning that we will take the positive branch of $\theta_{\pm}\mapsto\theta_{+}$.

For the first case, we will consider an initial wavefunction of Fock states. Then the evolution of these states is given by
\begin{equation}
    \left\langle\hat{n}_{a}(t)\right\rangle = n\left\vert\mathcal{F}_{1}(t)\right\vert^{2} + m\left\vert\mathcal{F}_{2}(t)\right\vert^{2}\qquad
    \left\langle\hat{n}_{a}(t)\right\rangle = m\left\vert\mathcal{G}_{1}(t)\right\vert^{2} + n\left\vert\mathcal{G}_{2}(t)\right\vert^{2},
\end{equation}
where $n, m\in\mathbb{N}$. For the second case, we put initial coherent states in each of the field, $\ket{\psi(0)} = \ket{\alpha, \beta, +} = \sum_{n,m}\ p_{n}(\alpha)p_{m}(\beta)\ket{n, m, +}$, where $p_{k}(\alpha)$ is the Poissonian photon distribution, defined as
\begin{equation}
    p_{k}(\alpha) = e^{-\frac{\vert\alpha\vert^{2}}{2}}\frac{\alpha^{k}}{\sqrt{k!}}.
\end{equation}
In this case, the relations for the evolution of coherent states in the initial field are 
\begin{equation}
    \langle\hat{n}_{a}(t)\rangle
    =
    \vert\alpha\vert^{2}\vert\mathcal{F}_{1}(t)\vert^{2}
    +
    \vert\beta\vert^{2}\vert\mathcal{F}_{2}(t)\vert^{2}
    +
    2\mathrm{Re}\!\left[\alpha\beta^{*}\mathcal{F}_{1}(t)\mathcal{F}_{2}^{*}(t)\right],
\end{equation}
\begin{equation}
    \langle\hat{n}_{b}(t)\rangle
    =
    \vert\beta\vert^{2}\vert\mathcal{G}_{1}(t)\vert^{2}
    +
    \vert\alpha\vert^{2}\vert\mathcal{G}_{2}(t)\vert^{2}
    +
    2\mathrm{Re}\!\left[\beta\alpha^{*}\mathcal{G}_{1}(t)\mathcal{G}_{2}^{*}(t)\right].
\end{equation}

These expressions illustrate how the average photon numbers in two output modes evolve when two coherent states, with amplitudes $\alpha$ and $\beta$, undergo a general linear optical transformation. Each output mode is a mixture of the original modes, with weights described by the time-dependent functions $\mathcal{F}_{j}(t)$ and $\mathcal{G}_{j}(t)$. Because coherent states behave like classical fields, the sums over Fock states collapse into simple intensity and interference terms. The terms proportional to $\vert\alpha\vert^{2}\vert\mathcal{F}_{1}\vert^{2}$ or $\vert\beta\vert^{2}\vert\mathcal{F}_2\vert^{2}$ represent the direct contribution of each input beam to the output intensity, corresponding to classical mode mixing. The cross terms, such as $\alpha\beta^{*}\mathcal{F}_{1}(t)\mathcal{F}_{2}^{*}(t)$, encode phase-sensitive interference between the two coherent inputs and survive because coherent states carry well-defined phases. Each expectation value therefore combines linear redistribution of intensities with an interference contribution controlled by the relative phase of $\alpha$ and $\beta$ and the dynamical mixing specified by the functions $\mathcal{F}_{j}(t)$ and $\mathcal{G}_{j}(t)$. The oscillation frequency is controlled by the normal-mode splitting $\vert\Omega_{A}^{(\pm)} - \Omega_{B}^{(\pm)}\vert$, which is of order $J$. Consequently, population transfer occurs on a time scale much longer than the bare Rabi period, a feature that is clearly visible in Fig.~\ref{fig:time_evolution}, where it displays the time evolution of the mean photon numbers $\langle \hat{n}_a(t) \rangle$ and $\langle \hat{n}_b(t) \rangle$ for representative parameters. Several pedagogically important features can be directly understood from the effective-Hamiltonian analysis. The oscillation period is set by $J$ and by the normal-mode splitting, not by $g_{a}$ or $g_{b}$. This reflects the elimination of fast atom--field exchange and the emergence of a renormalized, slow time scale. The out-of-phase behavior of $\langle\hat{n}_{a}(t)\rangle$ and $\langle\hat{n}_{b}(t)\rangle$ is the direct signature of the effective beam-splitter term $(\hat{a}^{\dagger}\hat{b} + \hat{a}\hat{b}^{\dagger})\hat{\sigma}_{z}$. Distinct curves correspond to branches $\pm 1$. Because $\Omega_{A}^{(\pm)}$, $\Omega_{A}^{(\pm)}$ and $\theta_\pm$ depend on these branches, the atom acts as a conditional controller of the field dynamics. When compared with simulations of the full Hamiltonian (solid curves), the effective description reproduces the slow envelope and phase of the oscillations, while small discrepancies arise from neglected higher-order terms. These deviations become visible only on long time scales, providing a clear diagnostic of the regime of validity. From a teaching standpoint, the figure encapsulates the central message of the paper: effective Hamiltonians do not merely approximate energies but reorganize the dynamics by introducing new emergent time scales that govern observable behavior.

\section{Conclusion}
\label{sec_cons}

The two-mode JCM provides an exceptionally clear platform for learning how effective Hamiltonians arise and how they can simplify the description of multimode light--matter interactions. In this work, we have developed a fully pedagogical derivation of the first-order effective Hamiltonian in the dispersive regime, beginning with a review of the single-mode JCM, extending the discussion to the two-mode case, and then applying a small-rotation (Schrieffer--Wolff) method to remove nonresonant excitation-exchange terms.

A central pedagogical insight is that once the rapidly oscillating atom--field exchange interactions are perturbatively eliminated, the remaining Hamiltonian reduces to a simple quadratic form that can be diagonalized by a geometric $\mathfrak{su}(2)$ rotation. This diagonalization reveals two normal modes whose frequencies depend on the atomic state, illustrating how virtual processes give rise not only to dispersive shifts but also to an effective beam-splitter interaction between the modes. The rotation angle structure, exemplified in Fig.~\ref{fig:theta_branches}, provides a geometric picture of the atom's role as a tunable mediator of mode hybridization.

By computing the time evolution of both Fock and coherent initial states, we highlighted the transparency of the effective-Hamiltonian approach: the dynamics reduce to independent harmonic evolution in the rotated basis, followed by a simple inverse rotation. This makes it easy to interpret phenomena such as conditional mode mixing, population exchange, and state-dependent beating patterns in terms familiar from classical optics, while preserving their quantum character. Such examples form a bridge between formal perturbative techniques and experimentally relevant behaviors in cavity QED, trapped-ion systems, and circuit QED platforms \cite{HarocheRaimond_Book, BlaisGambettaCQEDPed}.
 
From a teaching perspective, the two-mode dispersive JC model is an ideal laboratory for introducing effective Hamiltonians. It demonstrates how virtual processes alter observable dynamics, how $\mathfrak{su}(2)$ transformations naturally arise in multimode problems, and how diagonalization techniques can simplify time evolution. Furthermore, the method generalizes straightforwardly to more complex systems, including multimode cavities, arrays of resonators, and higher-dimensional atomic structures—making it a robust tool for students progressing into modern quantum optics and quantum information science.
 
We hope that the step-by-step exposition provided here offers both intuition and technique, equipping students and instructors with a clear foundation from which to explore more advanced applications of effective Hamiltonian methods in contemporary quantum platforms.

\section*{Acknowledgements}
The author would like to acknowledge the fruitful discussions with Dr. Jos\'e Récamier and M.Sc. Luis Medina at ICF. The author is also indebted to the ICF-UNAM for providing the physical and logistical facilities for the realization of this work.

\bibliographystyle{unsrt}
\bibliography{bib}

\end{document}